\documentclass[aps,twocolumn,amsmath,letterpaper]{revtex4}
\usepackage{amssymb}
\usepackage{graphicx}
\usepackage{array}
\usepackage{hhline}
\usepackage{longtable}

\begin{document}

    \title{Odd $\textbf{q}$-State Clock Spin-Glass Models in Three Dimensions,

     Asymmetric Phase Diagrams, and Multiple Algebraically Ordered Phases}

    \author{Efe Ilker$^{1}$ and A. Nihat Berker$^{1,2}$}
 \affiliation{$^1$Faculty of Engineering and Natural Sciences, Sabanc\i~University, Tuzla 34956, Istanbul, Turkey,}
    \affiliation{$^2$Department of Physics, Massachusetts Institute of Technology, Cambridge, Massachusetts 02139, U.S.A.}

\begin{abstract}

Distinctive orderings and phase diagram structures are found, from
renormalization-group theory, for odd $q$-state clock spin-glass
models in $d=3$ dimensions. These models exhibit asymmetric phase
diagrams, as is also the case for quantum Heisenberg spin-glass
models.  No finite-temperature spin-glass phase occurs. For all odd
$q\geqslant 5$, algebraically ordered antiferromagnetic phases
occur.  One such phase is dominant and occurs for all $q\geqslant
5$.  Other such phases occupy small low-temperature portions of the
phase diagrams and occur for $5 \leqslant q \leqslant 15$.   All
algebraically ordered phases have the same structure, determined by
an attractive finite-temperature sink fixed point where a dominant
and a subdominant pair states have the only non-zero Boltzmann
weights. The phase transition critical exponents quickly saturate to
the high $q$ value.


%05.10.Cc    Renormalization group methods
%64.60.ae    Renormalization-group theory
%64.60.De    Statistical mechanics of model systems
%75.10.Nr    Spin-glass and other random models
%75.50.Lk    Spin glasses and other random magnets

\end{abstract}

    \maketitle
    \def\s{\rule{0in}{0.28in}}
    \setlength{\LTcapwidth}{\columnwidth}

\section{Introduction}

Spin-glass problems \cite{NishimoriBook} continue to fascinate with
new orderings and phase diagrams under frustration \cite{Toulouse}
and ground-state entropy \cite{BerkerKadanoff,BerkerKadanoffCo}. The
extension of these models from the extensively studied Ising spin
models to less simple spins offer the possibility of completely new
orderings and phase diagrams.  We find that odd $q$-state clock
models are such cases.  Spins in odd $q$-state clock models cannot
be exactly anti-aligned with each other.  Furthermore, for a given
spin, its interacting neighbor has two states that give the
maximally misaligned pair configuration.  This fact immediately
injects ground-state entropy in the presence of antiferromagnetic
interactions, even without the frozen randomness of interactions of
the spin-glass system.

We have calculated, from renormalization-group theory, the phase
diagrams of arbitrary odd $q$-state clock spin-glass models in $d=3$
dimensions. We find that these models have asymmetric phase
diagrams, as is also the case for quantum Heisenberg spin-glass
models \cite{Heisenberg}.  They exhibit no finite-temperature
spin-glass phase.  For all odd $q\geqslant 5$, algebraically ordered
antiferromagnetic phases occur.  One such phase is dominant and
occurs for all $q\geqslant 5$.  Other such phases occupy a small
low-temperature portion of the phase diagram and occur for $5
\leqslant q \leqslant 15$.  All algebraically ordered phases have
the same structure, determined by an attractive finite-temperature
sink fixed point where a dominant and a subdominant pair states have
the only non-zero Boltzmann weights. The phase transition critical
exponents come from distinct critical fixed points, but quickly
saturate to the high $q$ value. Thus, a rich phase transition
structure is seen for odd $q$-state spin-glass models on a $d=3$
hierarchical lattice.

\section{The odd $\textbf{q}$-state clock spin-glass model and the renormalization-group method}
The $q$-state clock models are composed of unit spins that are
confined to a plane and that can only point along $q$ angularly
equidistant directions. Accordingly, the $q$-state clock spin-glass
model is defined by the Hamiltonian
    \begin{equation}
        \label{eq:1}
        \begin{split}
            -\beta \mathcal{H}=&\sum_{\langle ij \rangle}
            J_{ij}\vec{s}_i\cdot\vec{s}_j = \sum_{\langle ij \rangle}
            J_{ij}cos(\theta_{i}-\theta_{j}),
        \end{split}
    \end{equation}
where $\beta=1/k_{B}T$, at site $i$ the spin angle $\theta_{i}$
takes on the values $(2\pi/q)\sigma_i$ with
$\sigma_i=0,1,2,...,q-1$, and $\langle ij \rangle$ denotes that the
sum runs over all nearest-neighbor pairs of sites.  The bond
strengths $J_{ij}$ are $+J>0$ (ferromagnetic) with probability $1-p$
and $-J$ (antiferromagnetic) with probability $p$. This model
becomes the Ising model for q=2 and the XY model for
$q\rightarrow\infty$.

\begin{figure}[h!]
\centering
\includegraphics[scale=1]{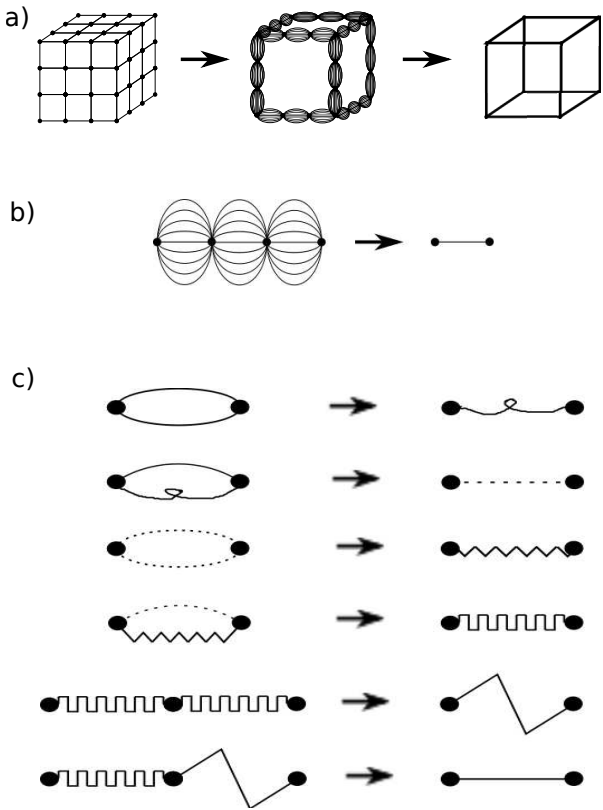}
\caption{(a) Migdal-Kadanoff approximate renormalization-group
transformation for the $d=3$ cubic lattice with the length-rescaling
factor of $b=3$. Bond-moving is followed by decimation. (b) Exact
renormalization-group transformation for the equivalent $d=3$
hierarchical lattice with the length-rescaling factor of $b=3$. (c)
Pairwise applications of the quenched probability convolution of
Eq.(5), leading to the exact transformation in (b).}\label{fig:1}
\end{figure}

The $q$-state clock spin-glass model, in $d=3$ dimensions, is
readily solved by a renormalization-group method that is approximate
on the cubic lattice \cite{Migdal,Kadanoff} and simultaneously exact
on the hierarchical lattice
\cite{BerkerOstlund,Kaufman1,Kaufman2,McKay,Hinczewski1}.
Hierarchical lattices have been used to study a variety of
spin-glass and other statistical mechanics
problems.\cite{Gingras2,Migliorini,Gingras1,Hinczewski,Guven,Ohzeki,Ozcelik,Gulpinar,Ilker1,Ilker2,Kaufman,
Barre, Monthus,
Zhang,Shrock,Xu,Herrmann1,Hermann2,Garel,Hartmann,Fortin,Wu,Timonin,Derrida,Thorpe,Hasegawa,Monthus2014,Lyra,
Xu2014, Hirose} Under rescaling, for $q>4$, the form of the
interaction as given in the rightmost side of Eq.(1) is not
conserved and one must therefore express the Hamiltonian more
generally, as
    \begin{equation}
        \label{eq:2}
        -\beta \mathcal{H}=\sum_{\langle ij
        \rangle}V(\theta_{i}-\theta_{j})\,.
    \end{equation}
The energy $V(\theta_{i}-\theta_{j})$ depends on the absolute value
of the angle difference, $|(\theta_{i}-\theta_{j})|$. Thus, the
renormalization-group flows are the flows of $q/2$ interaction
constants for even $q$ and the flows of $(q-1)/2$ interaction
constants for odd $q$. With no loss of generality, the maximum value
of $V(\theta_{i}-\theta_{j})$ is set to zero.

\begin{figure*}[]
\centering
\includegraphics*[scale=0.9]{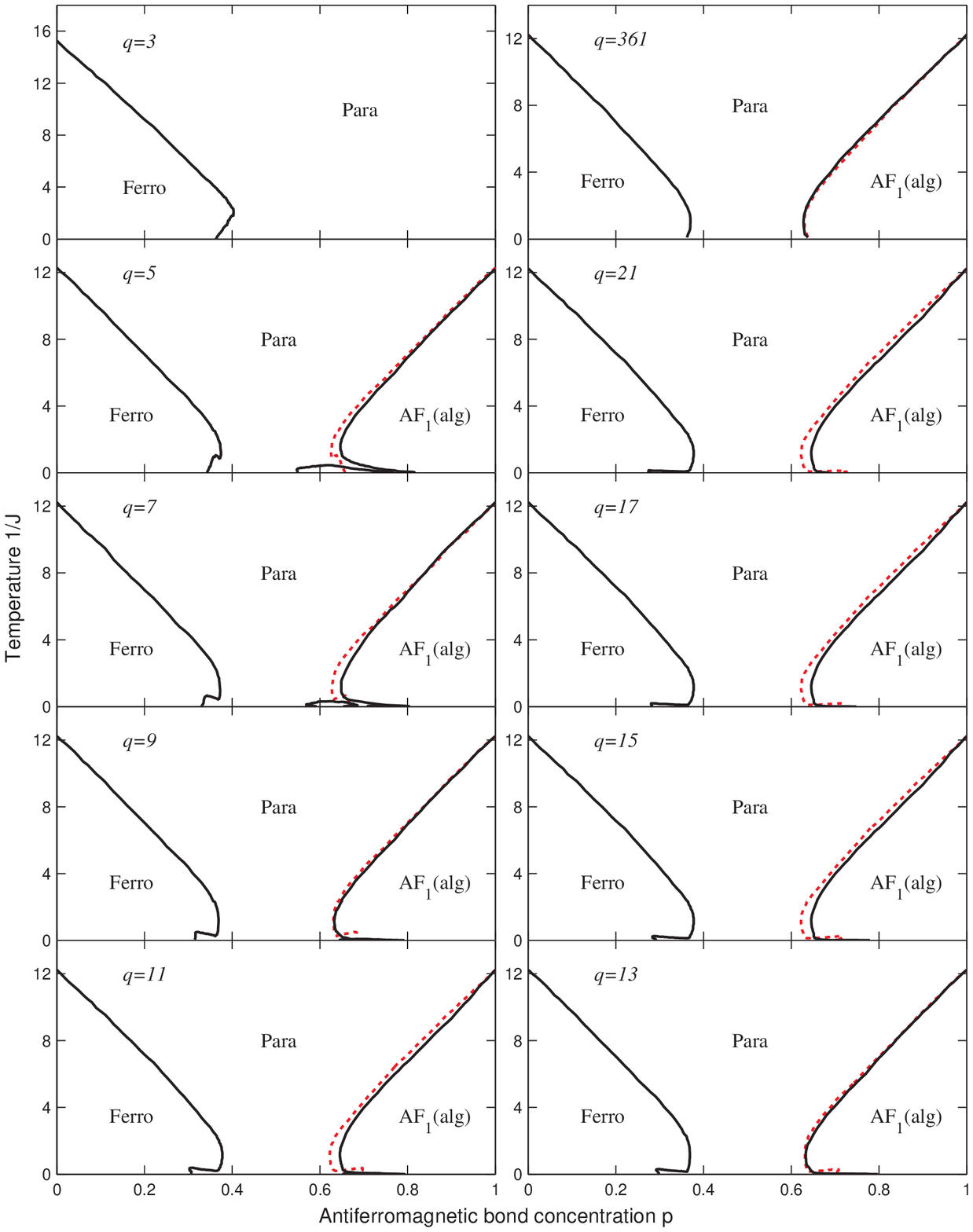}
\caption{(Color online) Calculated phase diagrams of the odd
$q$-state clock spin-glass models on the hierarchical lattice with
$d=3$ dimensions.  These phase diagrams do not have
ferromagnetic-antiferromagnetic symmetry, \textit{i.e.}, they are
not left-right symmetric with respect to the $p=0.5$ line. The phase
diagrams do not have a spin-glass phase, but show a multiplicity of
algebraically ordered phases on the antiferromagnetic side. The
phase diagrams show true reentrance (disordered-ordered-disordered)
as temperature is lowered at fixed antiferromagnetic bond
concentration $p$, on both the ferromagnetic and antiferromagnetic
sides of the phase diagram. The phase diagrams also show lateral,
true double reentrance
(ferromagnetic-disordered-ferromagnetic-disordered) as the
antiferromagnetic bond concentration $p$ is increased at fixed
temperature, only on the ferromagnetic side. No antiferromagnetic
ordering occurs for the lowest model, $q=3$. Algebracially ordered
antiferromagnetic phases occur for all higher $q \geq 5$ models.  In
these cases, the phase boundary between the dominant
antiferromagnetic algebraically ordered phase and the disordered
phase is slightly asymmetric with the phase boundary between the
ferromagnetic and disordered phases.  To make this slight asymmetry
evident, the latter boundary is also shown (dashed) reflected about
the $p=0.5$ line.  The lower temperature details of these phase
diagrams are shown in Fig.3}
\end{figure*}

\begin{figure*}[]
\centering
\includegraphics*[scale=0.9]{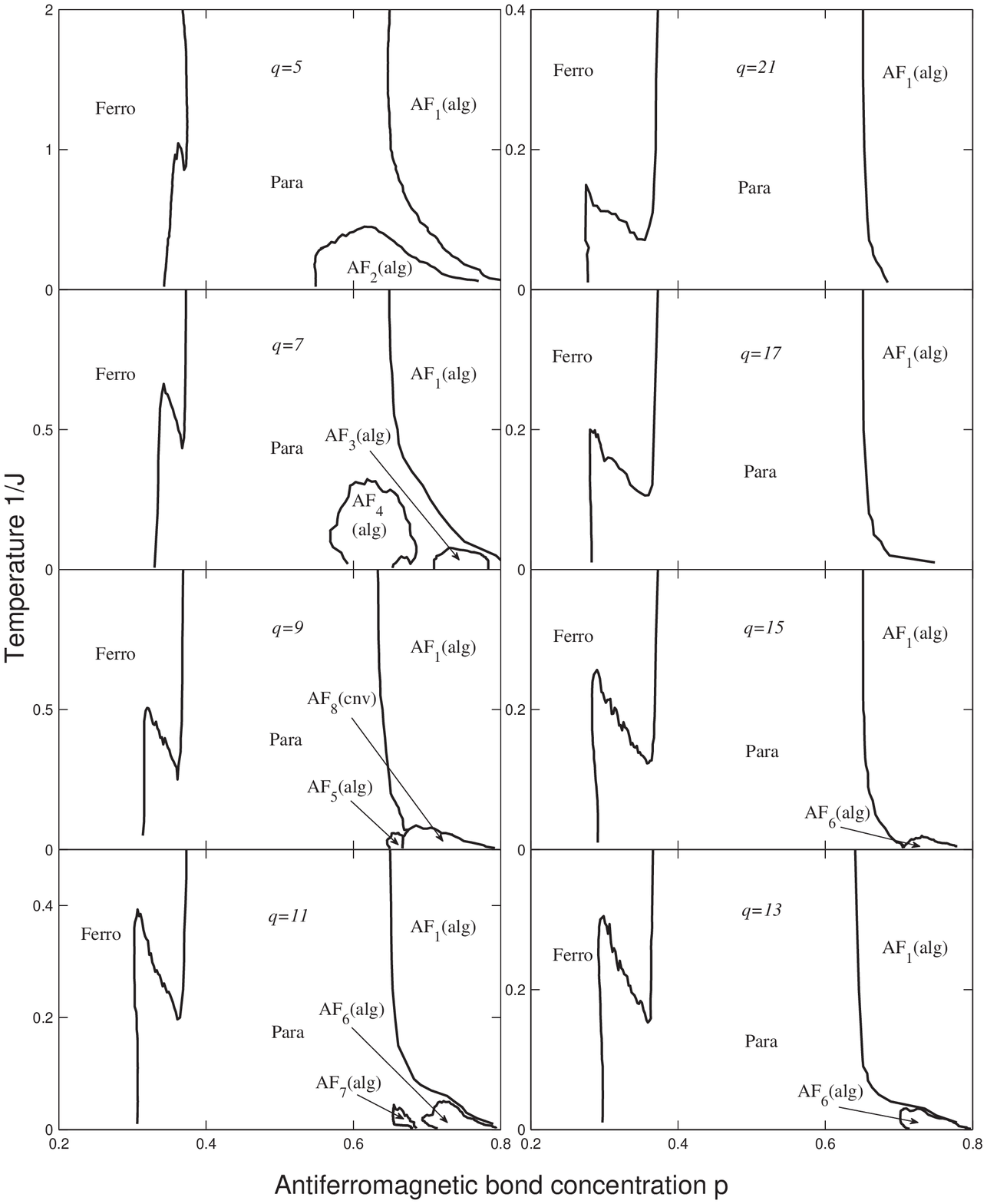}
\caption{Lower temperature details of the phase diagrams shown in
Fig.2}
\end{figure*}

The renormalization-group transformation, for spatial dimensions
$d=3$ and length rescaling factor $b = 3$ (necessary for treating
the ferromagnetic and antiferromagnetic correlations on equal
footing), is achieved by a sequence of bond moving
 \begin{equation}
        \label{eq:4}
        V_{bm}(\theta_1-\theta_2) + G_{12} = \sum_{n=1}^{b^{d-1}}
        V_n(\theta_1-\theta_2)
    \end{equation}
and decimation
\begin{equation}
        \label{eq:3}
        e^{V_{dec}(\theta_1-\theta_4)+G_{14}}=\sum_{\theta_2,\theta_3}
        e^{V_1(\theta_1-\theta_2)+V_2(\theta_2-\theta_3)+V_3(\theta_3-\theta_4)},
    \end{equation}
where the constants $G_{ij}$ are fixed by the requirement that the
maximum value of $V(\theta_{i}-\theta_{j})$ is zero.

The starting bimodal quenched probability distribution of the
interactions, characterized by $p$ and described above, is also not
conserved under rescaling. The renormalized quenched probability
distribution of the interactions is obtained by the convolution
\cite{Andelman}
\begin{multline}
\label{eq:5}
        P'(V'(\theta_{i'j'})) =\\
        \int{\left[\prod_{ij}^{i'j'}dV(\theta_{ij})P(V(\theta_{ij}))\right]}
         \delta(V'(\theta_{i'j'})-R(\left\{V(\theta_{ij})\right\})),
\end{multline}
where $R(\left\{V(\theta_{ij})\right\})$ represents the bond moving
and decimation given in Eqs.(3) and (4).  For numerical
practicality, the bond moving and decimation of Eqs.(3) and (4) are
achieved by a sequence of pairwise combination of interactions, as
shown in Fig.1(c), each pairwise combination leading to an
intermediate probability distribution resulting from a pairwise
convolution as in Eq.(5). We effect this procedure numerically, by
generating 5,000 interactions that embody the quenched probability
distribution resulting from each pairwise combination. Due to the
large number of phase diagrams (Figs. 2 and 3), a single realization
of quenched randomness is used. Each of the generated 5,000
interactions is determined by $(q-1)/2$ interaction constants. At
each pairwise convolution as in Eq.(5), 5,000 randomly chosen pairs
are matched by Eq.(3) or (4), and a new set of 5,000 is produced.

The different thermodynamic phases of the model are identified by
the different asymptotic renormalization-group flows of the quenched
probability distributions.  For all renormalization-group flows,
inside the phases and on the phase boundaries, Eq.(5) is iterated
until asymptotic behavior is reached. Thus, we are able to calculate
phase diagrams for any number of clock states $q$.  Similar previous
studies, on other spin-glass systems, are in Refs.
\cite{Gingras1,Migliorini,Gingras2,Hinczewski,Guven,Ohzeki,Ozcelik,Gulpinar,Ilker1,Ilker2}.

In a previous study \cite{Ilker1}, using the above method, we have
considered even values of $q$. In this study, we consider odd values
of $q$ and calculate the phase diagrams, which are not symmetric
around $p = 0.5$. For $q$ odd, the system does not have sublattice
spin-reversal $(\theta_i \rightarrow \theta_i + \pi)$ symmetry,
which leads to the asymmetric phase diagrams.  We obtain
qualitatively new features in the phase diagrams for odd $q$.  These
features do not occur for even $q$.

\section{Calculated phase diagrams for odd $\textbf{q}$-state clock spin glasses in $\textbf{d=3}$}

Our calculated phase diagrams for the odd
$q=3,5,7,9,11,13,15,17,21,361$-state clock spin-glass models are
shown in Fig. 2.  The lower temperature details of the phase
diagrams are given in Fig. 3.  All phase boundaries are second
order.

The phase diagrams of the odd $q$-state clock spin-glass models are
quite different from the even $q$ phase diagrams \cite{Ilker1}: The
odd $q$ phase diagrams do not have ferromagnetic-antiferromagnetic
symmetry, \textit{i.e.}, they are not left-right symmetric with
respect to the $p=0.5$ line. The odd $q$ phase diagrams do not have
a spin-glass phase, which is consistent with previous results
\cite{Gingras1,Ilker1} that the XY model, corresponding to the
$q\rightarrow\infty$ limit of the $q$-state clock models, does not
have a spin-glass phase on $d=3$ hierarchical lattices.  The odd $q$
phase diagrams show a multiplicity of algebraically ordered phases
(and one conventionally ordered phase) on the antiferromagnetic
side. All points in an algebraically ordered phase flow, under
renormalization group, to a single stable fixed point (sink) that
occurs at non-zero, non-infinite temperature. Convergence to this
stable critical fixed occurs, to 6 significant figures, within 5
renormalization-group transformations. Further convergence is
obtained for more renormalization-group transformations.  As seen in
Fig. 4, at each renormalization-group transformation, the quenched
probability distribution of interactions changes from the initial
(1-p) and p double-delta function, to eventually reach the critical
sink described below. Because of this flow structure, the
correlation length is infinite and the correlation function decays
as an inverse power of distance (as opposed to exponentially) at all
points in such an algebraically ordered phase. Such algebraically
ordered phases were previously seen by Berker and Kadanoff
\cite{BerkerKadanoff,BerkerKadanoffCo} for antiferromagnetic Potts
models and have since been extensively studied \cite{Qin, Saleur,
Redinz1, Redinz2, Jacobsen1, Jacobsen2, Ikhlef, Jacobsen3,
Jacobsen4}. The correlation function decay critical exponent has the
same value for all points in such a phase, since the
renormalization-group flows are to single fixed point, in contrast
to the continuously varying critical exponents in the algebraically
ordered phase of the $d=2$ XY model, where the flows are to a fixed
line.\cite{Kosterlitz,Jose,BerkerNelson}

The phase diagrams show true reentrance \cite{Gingras1}
(disordered-ordered-disordered) as temperature is lowered at fixed
antiferromagnetic bond concentration $p$, on both the ferromagnetic
and antiferromagnetic sides of the phase diagram. The phase diagrams
also show lateral, true double reentrance
(ferromagnetic-disordered-ferromagnetic-disordered) as the
antiferromagnetic bond concentration $p$ is increased at fixed
temperature, only on the ferromagnetic side.  Multiple reentrances
have previously been seen in liquid crystal
systems.\cite{Indekeu,Netz,Mazza}

No antiferromagnetic ordering occurs for the lowest model, $q=3$.
Algebracially ordered antiferromagnetic phases occur for all higher
$q \geq 5$ models.  In these cases, the phase boundary between the
dominant antiferromagnetic algebraically ordered phase and the
disordered phase is slightly asymmetric with the phase boundary
between the ferromagnetic phase and the disordered phase.  To make
this slight asymmetry evident, the latter boundary is also shown
(dashed) in Fig. 2 reflected about the $p=0.5$ line.  The phase
diagram for the XY model limit, namely odd $q\rightarrow\infty$, is
also shown in Fig. 2, calculated here with $q=361$ clock states. In
this limit, the distinction between odd and even $q$ disappears.
This suggests that the zero-temperature spin-glass phase
\cite{Grinstein} found for even $q\rightarrow\infty$ \cite{Ilker1}
also occurs for odd $q\rightarrow\infty$.

\begin{figure}[]
\centering
\includegraphics*[scale=1]{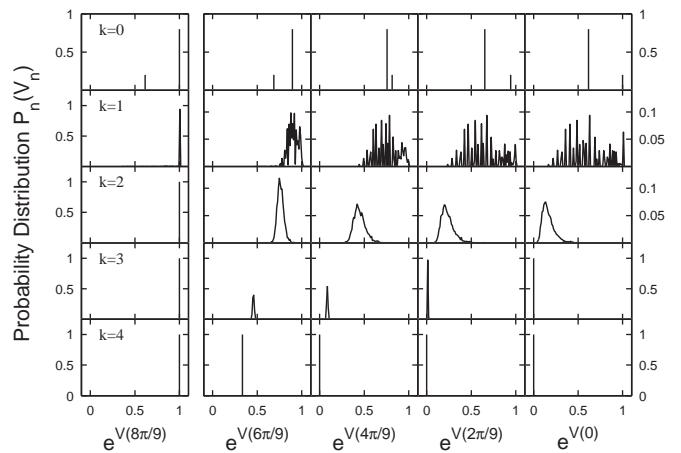}
\caption{Evolution of the quenched probability distribution under
successive renormalization-group transformations. The case of $q=9$,
starting with the initial condition temperature $1/J=4$ and
antiferromagnetic bond concentration $p=0.8$ is shown here.  For
$q=9$, the generalized interaction potential unavoidably generated
by the renormalization-group transformation is determined by 5
interaction constants (see Table I).  The renormalization-group
transformation gives the evolution, under scale change, of the
correlated quenched probability distribution
$P(V_0,V_1,V_2,V_3,V_4)$.  Shown in this figure are the projections
$P_0(V_0)=\int dV_1 dV_2 dV_3 dV_4 P(V_0,V_1,V_2,V_3,V_4)$ and
similarly for $P_1(V_1), P_2(V_2), P_3(V_3),$ and $P_4(V_4)$. Each
row corresponds to another renormalization-group step $k$, as marked
on the figure.  It is seen here that in four renormalization-group
transformations, the renormalized system essentially reaches the
critical phase sink described in Sec. IV:  The most misaligned pair
state is dominant with Boltzmann weight $e^{V(8 \pi /9)}=1$ and the
next-most misaligned pair state is also present but less dominant
with $e^{V(6 \pi /9)}=1/3$. The other two less misaligned pair
states and the aligned pair state have zero Boltzmann weight at the
sink. }
\end{figure}

\begin{figure}[]
\centering
\includegraphics*[scale=1]{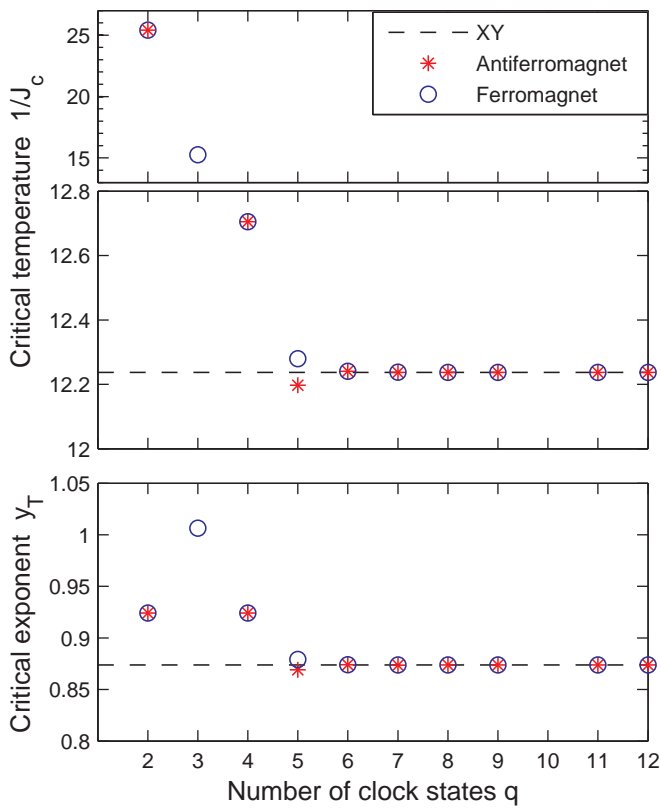}
\caption{(Color online) Top panel: Critical temperatures $1/J_C$ of
the ferromagnetic (circles) and antiferromagnetic (asterisks)
$q$-state clock models in $d=3$. Lower panel: Critical exponents
$y_T$ of the ferromagnetic (circles) and antiferromagnetic
(asterisks) $q$-state clock models in $d=3$.  In both panels, the
values exactly coincide for even $q$, due to the
ferromagnetic-antiferromagnetic symmetry that is present for even
$q$ but absent for odd $q$.}
\end{figure}

\section{Algebraically Ordered Phases, Finite-Temperature Renormalization-Group Sinks, and Ground-State Entropy}

Spins in odd $q$-state clock models cannot be exactly anti-aligned
with each other, \textit{i.e.}, $\theta_i - \theta_j = 2 \pi q_{ij}
/ q < \pi$, where $q_{ij}$ is an integer between 0 and $(q-1)/2$
inclusive. Furthermore, for a given spin, its interacting neighbor
has two states that give the maximally misaligned pair configuration
with $\theta_i - \theta_j = \pi (q-1)/q < \pi$.  Thus, for
antiferromagnetic interaction, this local degeneracy is of crucial
distinctive importance, injecting ground-state entropy into the
system, driving the sink of a would-be ordered phase to non-zero
temperature, and thereby causing algebraic order, as generally
explained in Ref. \cite{BerkerKadanoff,BerkerKadanoffCo}.

All points in the antiferromagnetic phases in the phase diagrams in
Figs. 2 and 3 flow under renormalization-group to $p=1$ (just as all
points in the one ferromagnetic phase flow to $p=0$). The most
extant antiferromagnetic phase in Fig. 2, labeled $AF_1(alg)$,
occurring for all odd $q\geqslant 5$ values, is an algebraically
ordered phase. All points in this phase flow to a completely stable
fixed point ("a phase sink" \cite{Berker0}) that is also a critical
point since it occurs at finite temperature
\cite{BerkerKadanoff,BerkerKadanoffCo}. Of the pair-interaction
Boltzmann weights $e^{V(\theta_i - \theta_j)}$, with $\theta_i -
\theta_j = \pi (q-1-2n)/q$, where $n=0$ is the most misaligned pair
state, $n=1$ is the next-most misaligned state, \textit{etc.}, until
$n=(q-1)/2$ is the completely aligned pair state, only two are
non-zero at this sink:  The most misaligned pair state, $n=0$, is
dominant with $e^{V(\pi (q-1)/q)}=1$ and the next-most misaligned
pair state, $n=1$, is also present but less dominant with $e^{V(\pi
(q-3)/q)}=1/3$. The other, less misaligned pair states, with $n
\geqslant 2$, and the aligned pair state have zero Boltzmann weight
at this sink. That these sink fixed-point Boltzmann weights are
applicable for all odd $q$ is consistent with the fact that the
$q-5$ less-misaligned pair states and one aligned pair state have
negligible Boltzmann weights at the sink fixed point, so that the
numerosity of $q$ does not matter. The finite difference between the
energies for $\theta_i - \theta_j = \pi (q-1)/q$ and $\theta_i -
\theta_j = \pi (q-3)/q$ establishes this sink as a
finite-temperature attractive critical fixed point. It can be shown
that, in the basin of attraction of a finite-temperature fixed
point, the order parameter is strictly zero, the correlation length
is infinite, and the correlations vanish algebraically with
distance.\cite{BerkerKadanoff,BerkerKadanoffCo,Fisher,McKay}

The evolution of the quenched probability distribution, under
successive renormalization-group transformations, towards such a
critical sink is shown in Fig. 4. The case of $q=9$, starting with
the initial condition temperature $1/J=4$ and antiferromagnetic bond
concentration $p=0.8$ is shown in the figure.  For $q=9$, the
generalized interaction potential unavoidably generated by the
renormalization-group transformation is determined by 5 interaction
constants (see Table I).  The renormalization-group transformation
gives the evolution, under scale change, of the correlated quenched
probability distribution $P(V_0,V_1,V_2,V_3,V_4)$.  Shown in Fig. 4
are the projections $P_0(V_0)=\int dV_1 dV_2 dV_3 dV_4
P(V_0,V_1,V_2,V_3,V_4)$ and similarly for $P_1(V_1), P_2(V_2),
P_3(V_3),$ and $P_4(V_4)$. Each row corresponds to another
renormalization-group step $k$, as marked on the figure.  It is seen
that in four renormalization-group transformations, the renormalized
system essentially reaches the critical phase sink described above:
The most misaligned pair state is dominant with Boltzmann weight
$e^{V(8 \pi /9)}=1$ and the next-most misaligned pair state is also
present but less dominant with $e^{V(6 \pi /9)}=1/3$. The other two
less misaligned pair states and the aligned pair state have zero
Boltzmann weight at the sink.

The less extant antiferromagnetic phases occur for specific $q$
values, at lower temperatures, and are disconnected from the most
extant antiferromagnetic phase $AF_1(alg)$. In $AF_2(alg)$, the two
sink Boltzmann weights have exchanged roles:  the next-most
misaligned pair state, $n=1$, is dominant with $e^{V(\pi
(q-3)/q)}=1$ and the most misaligned pair state, $n=0$, is also
present but less dominant with $e^{V(\pi (q-1)/q)}=1/3$.  In
$AF_3(alg)$, $AF_4(alg)$, $AF_5(alg)$, $AF_6(alg)$, $AF_7(alg)$,
these roles are played respectively by $n=2,0$, $n=1,2$, $n=2,1$,
$n=1,4$, $n=4,2$. On the other hand, $AF_8(cnv)$ is a conventionally
ordered phase, with a strong-coupling sink fixed point where $n=1$
and $n=4$ are equally dominant.

\newlength{\Oldarrayrulewidth}
\newcommand{\Cline}[2]{%
\noalign{\global\setlength{\Oldarrayrulewidth}{\arrayrulewidth}}%
  \noalign{\global\setlength{\arrayrulewidth}{#1}}\cline{#2}%
  \noalign{\global\setlength{\arrayrulewidth}{\Oldarrayrulewidth}}}

    \begin{table*}[h!]
        \begin{tabular}{! {\vrule width 1.7pt} c c c c c c c c c c c c c c c c ! {\vrule width 1.7 pt}  c c l l}

\Cline{1.1 pt}{1-16} \hline
V($\theta_{ij}$)  &\vline & n=0  &\vline &  n=1  &\vline &  n=2 &\vline &  n=3 &\vline & n=4 &\vline & n=5 &\vline &  n=6 & &  $y_T$ &\vline & relevant eigenvectors \\
\hline
q=5  &\vline & 0 &\vline &  -0.0905  &\vline &  -0.1502  &\vline & & & & & & & & & 0.869030&\vline & ( 1, 0.588 )\\
\hline
q=7 &\vline & 0  &\vline &  -0.0538 & \vline &  -0.1242 &\vline &  -0.1569 &\vline &  & & & & & & 0.873691 &\vline & ( 1, 0.782, 0.330 )\\
\hline
q=9  &\vline & 0   &\vline &  -0.0345 &\vline &  -0.0893  &\vline &  -0.1395 &\vline & -0.1599 &\vline &  & & & & 0.873709 &\vline & ( 1, 0.866, 0.544, 0.206 )\\
\hline
q=11  &\vline & 0   &\vline & -0.0238  &\vline &  -0.0649  &\vline &  -0.1111 &\vline &   -0.1475 &\vline & -0.1614 &\vline &   & &  0.873709 &\vline & ( 1, 0.909, 0.675, 0.387, 0.140 )\\
\hline
q=13 &\vline &  0  &\vline & -0.0173 &\vline & -0.0486 &\vline & -0.0873 &\vline & -0.1249  &\vline & -0.1523 &\vline &  -0.1623 & & 0.873709 &\vline & ( 1, 0.935, 0.759, 0.523, 0.287, 0.101 )\\
\hline
\Cline{1.1 pt}{1-16}
\end{tabular}
\caption{Antiferromagnetic critical fixed-point potentials
$V(\pi(q-1-2n)/q)$, critical exponents $y_T$, and corresponding
relevant eigenvectors of different odd $q$-state clock models. Thus,
each column progresses, from left to right, from the most misaligned
pair state $n=0$ to the aligned pair state $n=(q-1)/2$. For each
$q$, the relevant eigenvector is the (only) relevant eigenvector of
the $[(q-1)/2]$ \texttt{x} $[(q-1)/2]$ recursion matrix between the
independent $V(\theta_{ij})$. Although the fixed points and relevant
eigenvectors are distinct for different $q$, the critical exponents
quickly converge $y_T = 0.8737$.\\} \label{tab:1}
\end{table*}

It is thus seen that the stable sink fixed points that attract,
under renormalization-group flows, and characterize the
algebraically ordered phases have identical structure for all odd
$q\geqslant 5.$ A similar, but not identical, phenomenon occurs for
the unstable critical fixed points that control the
antiferromagnetic phase transitions. This is seen in Fig. 5 and
Table I, where the ferromagnetic $(p=0)$ and antiferromagnetic
$(p=1)$ critical temperatures $1/J_C$ are given as a function of
$q$. The fixed-point Boltzmann weight values $e^{V(\pi (q-1-2n)/q)}$
underpinning the antiferromagnetic phase transitions, as well as the
critical exponents $y_T$ and corresponding relevant eigenvectors are
given for different $q$ in Table I. For each $q$, the relevant
eigenvector is the (only) relevant eigenvector of the $[(q-1)/2]$
\texttt{x} $[(q-1)/2]$ recursion matrix between the independent
$V(\theta_{ij})$. Although the fixed points and relevant
eigenvectors are distinct for different $q$, the critical
temperatures and critical exponents quickly converge, for high $q$,
to $1/J_C = 12.2373$ and $y_T = 0.8737$. The critical temperatures
and exponents thus show differences for low q. The convergence for
high q of the critical temperatures at p=0 and p=1 is expected,
since the q-state clock models approach the XY model for large q,
with identical antiferromagnetic and ferromagnetic behavior.

\section{Conclusion}

We have calculated, from renormalization-group theory, the phase
diagrams of arbitrary odd $q$-state clock spin-glass models in
$d=3$. These models have asymmetric phase diagrams, as is also the
case for quantum Heisenberg spin-glass models \cite{Heisenberg}. For
all odd $q\geqslant 5$, algebraically ordered antiferromagnetic
phases occur.  One such phase is dominant and occurs for all
$q\geqslant 5$.  Other such phases occupy small low-temperature
portions of the phase diagrams and occur for $5 \leqslant q
\leqslant 15$.  All algebraically ordered phases have the same
structure, determined by an attractive finite-temperature sink fixed
point where a dominant and a subdominant pair states are the
non-zero Boltzmann weights.  The phase transition critical
exponents, on the other hand, vary with $q$ only at low $q$.

A rich and distinctive phase transition structure is thus seen for
odd $q$-state spin-glass models on a $d=3$ dimensional hierarchical
lattice.

\begin{acknowledgments}
Support by the Alexander von Humboldt Foundation, the Scientific and
Technological Research Council of Turkey (T\"UBITAK), and the
Academy of Sciences of Turkey (T\"UBA) is gratefully acknowledged.
\end{acknowledgments}

\end{document}